\documentclass[a4paper]{article}

\usepackage{INTERSPEECH2022}

\usepackage{graphicx}
\usepackage[utf8]{inputenc} 
\usepackage[T1]{fontenc}    
\usepackage{url}            
\usepackage{booktabs}       
\usepackage{amsfonts}       
\usepackage{nicefrac}       
\usepackage{microtype}      
\usepackage[dvipsnames]{xcolor}         
\usepackage{todonotes}
\usepackage{tabularx}
\usepackage{float}
\usepackage[square,sort,comma,numbers]{natbib}
\usepackage{xcolor}

\usepackage{hyperref}       
\hypersetup{
    colorlinks,
    linkcolor={red!80!black},
    citecolor={blue!50!black},
    urlcolor={blue!80!black}
}

\definecolor{colora}{RGB}{216,27,96}
\definecolor{colorb}{RGB}{30,136,229}
\definecolor{colorc}{RGB}{255,193,7}
\definecolor{colord}{RGB}{0,77,64}

\title{Content-Context Factorized Representations for Automated Speech Recognition}
\name{David M. Chan$^{1,2,*}$\thanks{${^{1}}$ Work done during an internship at Amazon Alexa AI}\thanks{${^{*}}$ Corresponding Author}, Shalini Ghosh$^2$}
\address{
  $^1$University of California, Berkeley (EECS)\\
  $^2$Amazon Alexa AI}
\email{davidchan@berkeley.edu, ghosha@amazon.com}

\begin{document}

\maketitle

\begin{abstract}
  Deep neural networks have largely demonstrated their ability to perform automated speech recognition (ASR) by extracting meaningful features from input audio frames. Such features, however, may consist not only of information about the spoken language content, but also may contain information about unnecessary contexts such as background noise and sounds or speaker identity, accent, or protected attributes. Such information can directly harm generalization performance, by introducing spurious correlations between the spoken words and the context in which such words were spoken. In this work, we introduce an unsupervised, encoder-agnostic method for factoring speech-encoder representations into explicit content-encoding representations and spurious context-encoding representations. By doing so, we demonstrate improved performance on standard ASR benchmarks, as well as improved performance in both real-world and artificially noisy ASR scenarios.
\end{abstract}
\vspace{0.5em}
\section{Introduction \& Background}

Despite recent progress in methods for automated speech recognition (ASR), understanding speech in noisy situations remains a challenging problem, even in high resource languages such as English. Not only must speech encoders understand and model the utterance's natural language, but they must also model a wide range of background and speaker-specific phenomena. Recent work \cite{shi2022learning, chan2021multi, bertoin2021disentanglement, kwon2020intra} have shown that explicitly modeling the context of a training sample, such as visual context~\cite{chan2021multi, shi2022learning}, the background of an image~\cite{bertoin2021disentanglement}, the speaker~\cite{kwon2020intra, du2021identity}, and even the dataset identity~\cite{chen2021scenario}, can help to improve modeling performance. 

\begin{figure}[t]
    \centering
    \includegraphics[width=\linewidth]{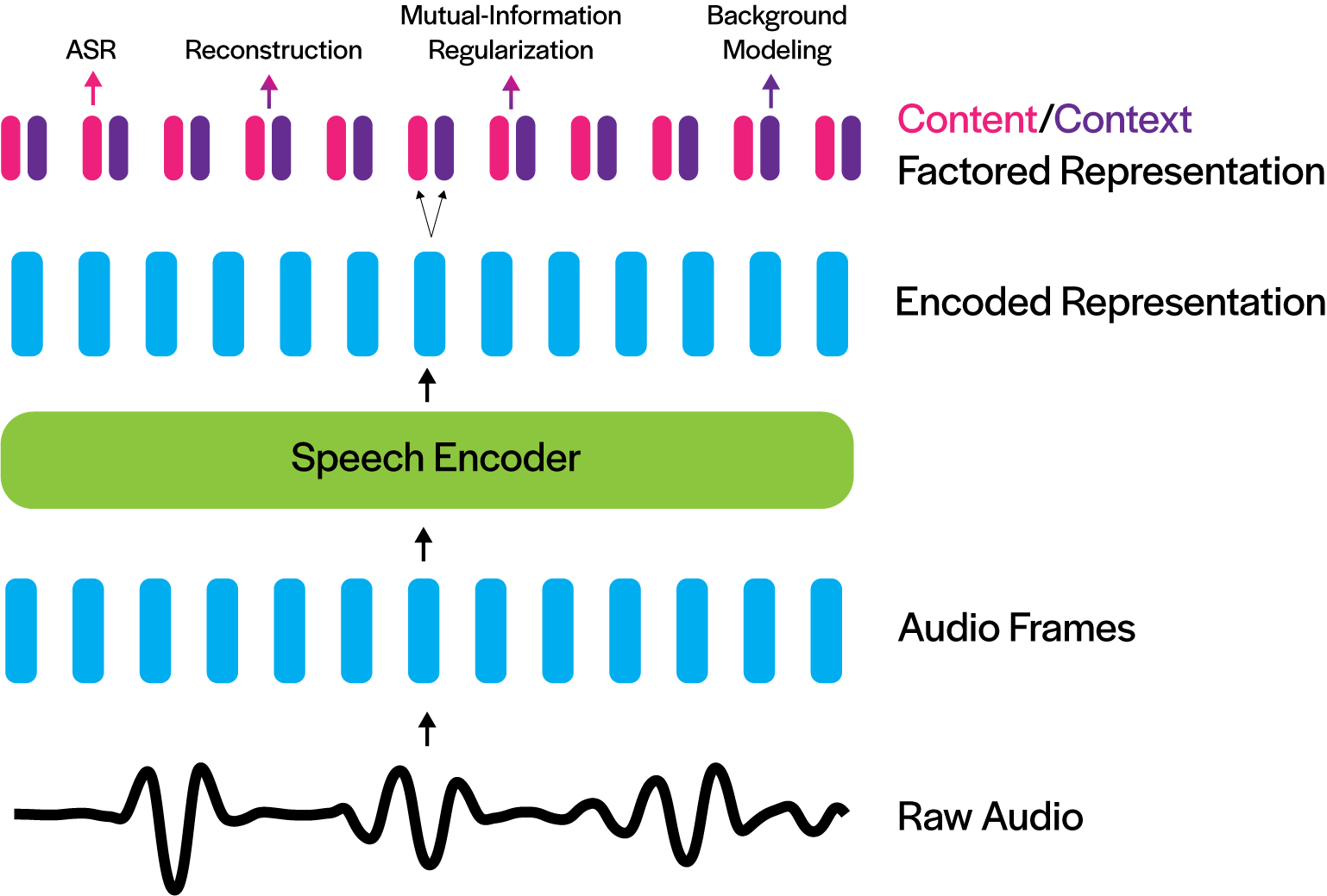}
    \caption{An overview of our factored content-context factored model. We first encode raw audio using an arbitrary frame-centric speech encoder, and then learn a content-context disentangled representation based on the encoded frames, which is adjusted through several regularization losses including background modeling, labeled ASR, reconstruction (and masked language modeling). To force disentangling of the latent space, we minimize the mutual information between the two representations, which should help improve robustness through the elimination of spurious context-content correlations.}
    \label{fig:teaser}
\end{figure}

Research leveraging factored representations to disentangle latent variables has largely been limited to the vision domain. Pioneered by Tenenbaum et al. \cite{tenenbaum2000separating}, who used bi-linear models to separate style from content, such methods have repeatedly shown \cite{rifai2012disentangling, mathieu2016disentangling, chen2016infogan, higgins2016beta, sanchez2020learning}  that some form of latent-factor disentanglement can significantly improve the generalization performance of vision models. However, such methods have largely remained absent from work in the speech domain, particularly in the field of automated speech recognition. Recently, Xin et al.~\cite{xin2021disentangled} and Gong et al.~\cite{gong2021using} demonstrated the benefits of disentangled representations when generating speech from text, Wang et al.~\cite{wang2021vqmivc} showed the same for voice conversation, while Kwon et al.~\cite{kwon2020intra} demonstrated gains in speaker recognition. Thus, it is clear that explicitly disentangled representations have the potential to improve generalization in a wide range of speech-related problems. 

In this work, we aim to extend the research in disentangled representation to the field of automated speech recognition. We argue that any speech utterance can be factored along two major axes: content and context. The content of an utterance is the raw natural language contained within the data, while the context of an utterance is a representation of the background environment, environmental cues, speaker identity, and other spurious content. By disentangling these representations, we aim to improve the generalization of the model, as spurious correlations will be encoded in the context representation over the content representation (which will be used for downstream ASR). 

\begin{figure*}[t]
    \centering
    \includegraphics[width=\linewidth]{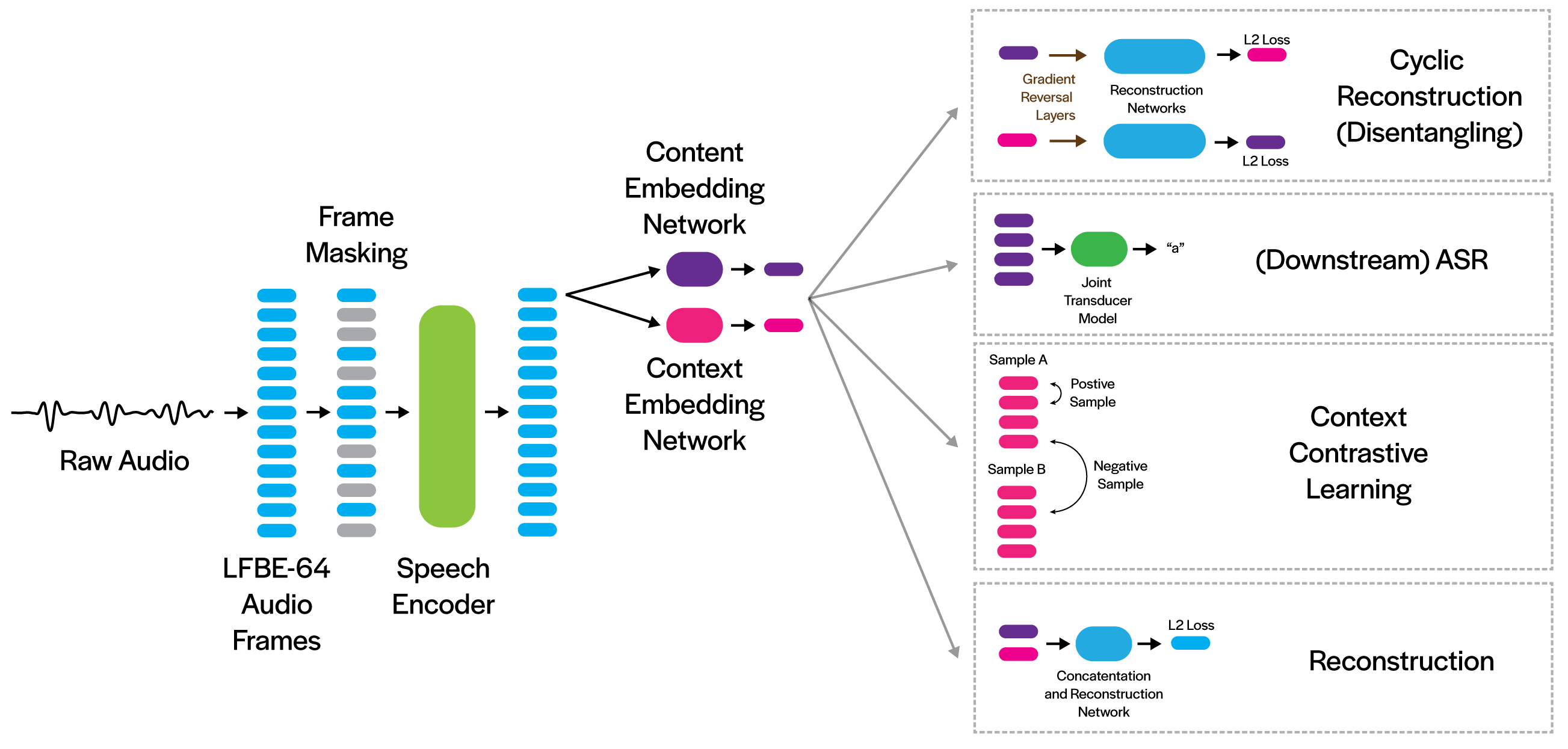}
    \caption{Overview of our proposed architecture. Raw audio features are extracted and then masked in a BERT-style masking approach to preserve noise robustness \cite{devlin2018bert}. We then pass the masked frames through an arbitrary speech encoder, and from those embeddings, generate a content embedding (containing the content of the frame) and a context embedding (containing background and noise information). Using these embeddings, we perform several downstream tasks to improve and separate the representations, as discussed in \autoref{sec:methods}.}
    \label{fig:model}
\end{figure*}

This work consists of several primary contributions:
\begin{enumerate}
    \item We introduce the first-ever ASR-centric and encoder-agnostic model approach for factorizing the latent space of a speech encoder into content-centric and context-centric representations.
    \item We develop several unsupervised loss functions based on contrastive learning, masked language modeling, and mutual information optimization, for training factored representations without additional labeled speech data.
    \item We demonstrate that latent-factored representations show improved performance on standard ASR data, as well as significantly improved performance in both artificially noisy and real-world noisy scenarios.
\end{enumerate}

\section{Methods}
\label{sec:methods}

In this work, we aim to factor the latent representation of an arbitrary frame-centric speech encoder into two components: a content-centric representation useful for ASR, and a context-centric component that encodes background and spurious correlations. Our goals in designing such a content-context disentangled representation are several-fold:
\begin{enumerate}
    \item Maximize the mutual information between the content representation, and the correct ASR labels.
    \item Encourage the context representation to encode background features and noise.
    \item Minimize the mutual information between the context-representation and the content representation.
\end{enumerate}

In this work, we achieve the second and third goals in an unsupervised fashion. In the future, we would like to explore options for achieving the first goal (i.e., maximizing the mutual information between the content-representation and correct phonemes) in an unsupervised way as well. 
Our overall architecture is specified in \autoref{fig:model}, and consists of two main components: (a) a factor-generator network that splits the frame-level outputs of an existing speech encoder into several factored representations of dimension $F$, and (b) a set of losses which encourage the above goals. 

\subsection{Factorized Learning with Cyclic Reconstruction}
\label{subsec:factorized}

Unfortunately, many methods \cite{rifai2012disentangling, mathieu2016disentangling} for disentangled representations require access to additional labeled data to supervise the paired factors. With a lack of labeled data already available for under-resourced languages, such methods are out of reach in most ASR application domains. We focus on learning disentangled representations in an unsupervised manner, e.g., unsupervised disentangling of features through the minimization of mutual information between latent representations during training. Minimizing the mutual information between two vectors is notably difficult, as no explicit upper bound for the mutual information can be directly optimized. Most recent work on learning disentangled representations such as Sanchez et. al. \cite{sanchez2020learning} has relied on discriminator-centric estimation methods such as MINE \cite{belghazi2018mine}, which provides an approach for estimating the mutual information using a GAN-like discriminator. A significant drawback of these methods is instability during training caused by the adversarial losses, compounded with the complexity of min-max learning in a multi-loss framework.

Thus, instead of leveraging a discriminative model, we minimize the mutual information between our factors using DiCyR \cite{bertoin2021disentanglement}, which leverages cyclic-reconstruction and gradient-reversal layers \cite{ganin2016domain} to maintain the constraint that the content representation shares little mutual information with the context representation. Let $\Pi_{\text{context}}(e_i)$ be the context encoding of the frame $e_i$, and $\Pi_{\text{content}}(e_i)$ be the content encoding of speech-encoded representation $e_i$ (encoding frame $f_i$). Leveraging DiCyR, we also define three models, $\phi_{\text{content}}$, $\phi_{\text{context}}$ and $\phi_{\text{joint}}$ which are designed to reconstruct the content from the context, context from the content, and original audio frames from both context and content respectively. Thus, we optimize:
\begin{equation}
    L_{\text{mi}} = L_{\text{m-content}} +  L_{\text{m-context}} + L_{\text{m-joint}}
\end{equation}
where
\begin{align*}
    L_{\text{m-content}} = ||\Pi_{\text{content}}(e_i) - \phi_{\text{content}}(GRL(\Pi_{\text{context}}(e_i)))||_2^2 \\
    L_{\text{m-context}} = ||\Pi_{\text{context}}(e_i) - \phi_{\text{context}}(GRL(\Pi_{\text{content}}(e_i)))||_2^2 \\
    L_{\text{m-joint}} = ||f_i - \phi_{\text{joint}}(\Pi_{\text{content}}(e_i), \Pi_{\text{context}}(e_i)))||_2^2 \\
\end{align*}
and $GRL(\cdot)$ is a gradient reversal layer. In practice, both the $\phi$ and $\Pi$ projections are implemented as shallow MLPs with several hidden layers and ReLU activation.

\subsection{Unsupervised Background-Contrastive Learning}
\label{subsec:bgcl}

Minimizing the mutual information between the content and context representations may be insufficient alone to improve the quality of the model, as context-specific (but rare) information can still propagate through the content representation. To reduce this likelihood, we explicitly introduce the notion that the context-representation should fully model the background information, by encouraging the context-representation to be invariant between frames. To facilitate learning this invariant, we leverage a contrastive learning framework based on SimCLR \cite{chen2020simple}, which has shown promise when modeling invariants in the Computer Vision domain. A similar technique was used in Baevski et al. \cite{baevski2020wav2vec} in the ASR domain. Our implementation exploits the idea that within an utterance, the context embedding should remain the same (while the content often differs). Let $\Pi_{\text{context}}(e^k_i)$ be the speech-encoded context representation of sample $k$, frame $i$. Thus, we can define the loss for a sample as:
\begin{equation*}
    l_{k, i, j} = - \log{\frac{exp(\Pi_{\text{context}}(e^k_i) \cdot \Pi_{\text{context}}(e^k_j))
    }{\sum_{l=1}^{M} exp(\Pi_{\text{context}}(e^k_i) \cdot \Pi_{\text{context}}(e^l_{min(j,|l|)}))}}
\end{equation*}
\begin{equation}
    L_{\text{contrast}} = \frac{1}{N} \sum_{k=1}^N \frac{1}{|k|^2}\sum_{i=1}^{|k|}\sum_{j=1}^{|k|} l_{k,i,j}
\end{equation}
where $|k|$ is the length of sample $|k|$, $N$ is the batch size, and $M$ is the number of sampled negatives.

\subsection{Additional Noise-Robustness with Masked Language Modeling}
\label{subsec:mlm}

Recently, HuBERT \cite{hsu2021hubert} has demonstrated that pre-training speech encoders using masked language modeling techniques, drawn from the NLP domain, can significantly improve the robustness of ASR models, particularly in noisy domains. As shown in \autoref{fig:model}, we modify the input to our base speech encoder to perform frame-level masking in addition to SpecAugment masking, masking each frame with a probability $\pi_{\text{mask}}$. In most of our experiments, we find that setting $\pi_{\text{mask}} = 0.15$ gives the best performance.

\subsection{Experimental Design}

\noindent\textbf{Speech Encoders:} Because our method is speech encoder agnostic, we evaluate our method using two speech encoders: An RNN-based encoder similar to the one in use in RNN-T \cite{graves2013speech} encoder, and the more recent Conformer encoder \cite{gulati2020conformer}, which leverages attention-based modeling. For the RNN-T encoder, we use a six layer LSTM encoder with a hidden dimension of 1024 (66M Params). For the conformer encoder on Librispeech, we follow Gulati et al. \cite{gulati2020conformer} and use a 17 layer encoder with a hidden dimension of 512 (118M Params). For internal Alexa-AI data, we use a conformer model with 208.37M parameters. For the ASR models, we use an LSTM-based decoder with two layers, and a hidden dimension of 1024. All models leverage ReLU activations, batch-normalization, and dropout of $0.1$. For the ASR tokenization, we use a sentence-piece model \cite{kudo2018sentencepiece} with a vocab size of 640 (librispeech) and 4096 (internal). \\

\noindent\textbf{Losses:} Combining the losses from subsections \ref{subsec:factorized}, \ref{subsec:bgcl}, and \ref{subsec:mlm}, in addition to the standard joint-model ASR loss \cite{graves2013speech} using the content embeddings as the speech encoder output (as shown in \autoref{fig:model}, which we refer to as $L_{\text{asr}}$) we optimize the combined loss function:
\begin{equation}
    L_{\text{total}} = L_{\text{asr}} + \lambda_{\text{mi}}L_{\text{mi}} + \lambda_{\text{contrast}}L_{\text{contrast}}
\end{equation}
where the $\lambda$ values represent hyper-parameter tradeoffs between the different loss terms.

In practice, this loss function has heavy regularization effects. To balance these effects while retaining the same number of parameters, we find that decreasing the dropout ratio and learning rate during training can lead to improved performance. \\

\noindent\textbf{Datasets:} We both train and evaluate the proposed model on the LibriSpeech \cite{panayotov2015librispeech} dataset, which consists of 970 hours of labeled speech. Librispeech consists of two splits: ``test-clean" and ``test-other". ``Test-clean" is composed of 5.4 hours of relatively clean speech, while ``test-other" consists of 5.1 hours of uncleaned speech data.  In addition to Librispeech, we demonstrate the real-world performance of our method by training on 300K hours of transcribed and self-distilled unlabeled Alexa speech data. Using this model, we present evaluation results on several Alexa-AI datasets: ``Base" containing 41K utterances that well-represent general Alexa speech tasks, and ``Rare", containing 48K utterances representing the long-tailed distribution of rare words. For training on the Alexa data, we use 120K hours of labeled utterances, with an additional 300K hours of self-supervised training data. All internal datasets have been processed to de-identify user data. \\

\noindent\textbf{Data Pre-Processing:} Speech utterances are pre-processed by first performing a global normalization, where audio signals are normalized to have zero mean, and unit standard deviation. Following normalization we extract, as input, 64-dimensional Log-Filterbank Energy (LFBE) features  \cite{oppenheim2001discrete}. Three frames are then stacked to generate 192-dimensional inputs, which are then augmented using SpecAugment \cite{park2019specaugment} with two frequency masks, twenty time masks, an adaptive multiplicity of $0.04$, a maximum ratio of masked frequencies of $27/80$ and a maximum ratio of masked time frames of $0.04$. \\

\noindent\textbf{Artificial Noise:} To simulate noisy situations, we perform data-mixing during testing where two utterances $u_a$ and $u_b$ are mixed at a frame level to generate $u_m = (1 - \alpha_m) u_a + \alpha_m u_b$ (Note that both utterances are magnitude normalized a-priori, and $\mu_b$ is clipped to the length of $\mu_a$). The label of $\mu_a$ is preserved as a reference label for $\mu_m$. Utterances are mixed at random, however, the chosen mixed utterances are the same across evaluations.  \\

\noindent\textbf{Implementation and Optimization Details:} In our experiments, we use a factor-dimension $F$ of $512$, and a factor-projection model consisting of three hidden layers, ReLU activations, and a hidden dimension of $512$.  In practice, we found that for masked language modeling, a masking ratio of $0.15$ was the best during training. When trading off between features, we found that $\lambda_{contrast}=0.3$, and $\lambda_{mi}=0.1$ to be optimal on a dev-set. For Librispeech, the model is implemented in Tensorflow, and is trained using 24 Nvidia V-100GPUs for 65 epochs with a batch size of 2048 and the Adam optimizer, with a learning rate of $3e^{-4}$. For the Alexa-AI datasets, the model is trained using 104 Nvidia V-100 GPUs for $120$ epochs with a batch size of $832$ and the Adam optimizer, with a warmup/hold learning rate schedule with $10,000$ warmup steps and a maximum learning rate of $5e^{-3}$. 

\section{Results \& Discussion}

\begin{table}
\footnotesize
\caption{\small Results summary of word error rate for the Librispeech dataset with no additional language model. FR: Factored Representation, BG-C: Background-Contrastive Objective, MLM: Masked Language Modeling}\label{tab:r1}
\begin{tabularx}{\linewidth}{lrrrrr}
	\toprule
		\textbf{Method} & FR & BG-C & MLM & \textbf{test-clean} & \textbf{test-other} \\
		
	\midrule
	\textbf{RNN-T} &&&&&\\
	    & \multicolumn{3}{r}{{\it\footnotesize Baseline}}          & 6.05  & 15.43  \\
	    & \checkmark &            &            & 6.06 \tiny{(+0.01\%)} &  15.42 \tiny{(-0.06\%)}  \\
	    &            & \checkmark &            & 6.08 \tiny{(+0.05\%)} & 15.46 \tiny{(+0.19\%)}  \\
	    &            &            & \checkmark & 5.98 \tiny{(-1.16\%)} & 15.07  \tiny{(-2.33\%)}  \\
	    & \checkmark & \checkmark &            & 5.91 \tiny{(-2.31\%)}& 15.01 \tiny{(-2.72\%)}  \\
	    & \checkmark &            & \checkmark & 5.87 \tiny{(-2.97\%)}& 14.84 \tiny{(-3.82\%)} \\
	    & \checkmark & \checkmark & \checkmark & \textbf{5.80} \tiny{(-4.13\%)} & \textbf{14.61} \tiny{(-5.31\%)}  \\
	 \midrule
	 \textbf{Conformer} &&&&& \\
	  & \multicolumn{3}{r}{{\it\footnotesize Baseline}} & 2.13 & 4.31  \\
	    & \checkmark &            &            & 2.12 \tiny{(-0.47\%)} & 4.28 \tiny{(-0.70\%)}  \\
	    &            & \checkmark &            & 2.14 \tiny{(+0.47\%)}  & 4.30 \tiny{(-0.23\%)}  \\
	    &            &            & \checkmark & 2.10 \tiny{(-1.41\%)}  & 4.26 \tiny{(-1.16\%)}  \\
	    & \checkmark & \checkmark &            & 2.11 \tiny{(-0.90\%)}  & 4.20 \tiny{(-2.55\%)}  \\
	    & \checkmark &            & \checkmark & 2.09 \tiny{(-1.87\%)}  & 4.16 \tiny{(-3.48\%)}  \\
	    & \checkmark & \checkmark & \checkmark & \textbf{2.07} \tiny{(-2.81\%)}  & \textbf{4.10} \tiny{(-4.77\%)} \\
	\bottomrule
\end{tabularx}
\end{table}
\begin{table}
\small
\caption{\small Relative percent improvements of our proposed models over baseline models (no language model) on the Alexa-AI Base and Alexa-AI Rare datasets.}\label{tab:r2}
\begin{tabularx}{\linewidth}{Xrrrr}
	\toprule
		\textbf{Method} & {\tiny rel.} WER & {\tiny rel.} SUB & {\tiny rel.} INS & {\tiny rel.} DEL \\
	\midrule
	\textbf{Alexa-AI Base} &&&& \\
	$\quad$ RNN-T  & 2.1\% & 3.0\% & 2.1\% & 5.1\% \\
	$\quad$ Conformer & 4.47\% & 4.3\% & 0.01\% & 7.5\% \\
	\textbf{Alexa-AI Rare} &&&& \\
	$\quad$ RNN-T & 3.9\% & 2.7\% & 0.06\% & 1.9\% \\
	$\quad$ Conformer & 4.50\% & 4.9\% & 5.49\% & 2.9\%  \\
	\bottomrule
\end{tabularx}
\vspace{-1.5em}
\end{table}
\begin{table}
\footnotesize
\caption{\small Word Error Rate of our proposed models vs. baseline models for test sets of the Librispeech dataset with artificially noisy test data, generated by sample mixing with rate $\alpha_m$. }\label{tab:r4}
\begin{tabularx}{\linewidth}{llll}
	\toprule
		\textbf{Noise} & \textbf{Method}  & \textbf{test-clean} & \textbf{test-other} \\
	\midrule
	\textbf{$\alpha_m=0$} &&&   \\
	    & RNN-T             & 6.05 & 15.43  \\
	    & RNN-T (Ours)      & \textbf{5.80} \tiny{(-4.13\%)} & \textbf{14.61} \tiny{(-5.31\%)}  \\
	    & Conformer         & 2.13 & 4.31  \\
	    & Conformer (Ours)  & \textbf{2.07} \tiny{(-2.81\%)}  & \textbf{4.10} \tiny{(-4.77\%)}  \\
	 \midrule
	 \textbf{$\alpha_m=0.1$} &&&   \\
	    & RNN-T             & 6.19  & 15.88   \\
	    & RNN-T (Ours)      & \textbf{5.82} \tiny{(-3.83\%)}  & \textbf{15.13} \tiny{(-4.68\%)}   \\
	    & Conformer         & 2.21  & 4.53   \\
	    & Conformer (Ours)  & \textbf{2.12}  \tiny{(-4.07\%)} & \textbf{4.29} \tiny{(-5.30\%)}  \\
	 \midrule
	 \textbf{$\alpha_m=0.25$} &&&   \\
	    & RNN-T             & 7.58  & 20.01   \\
	    & RNN-T (Ours)      & \textbf{7.25} \tiny{(-4.35\%)}  & \textbf{18.99} \tiny{(-5.10\%)}  \\
	    & Conformer         & 2.67 & 5.60   \\
	    & Conformer (Ours)  & \textbf{2.54} \tiny{(-4.86\%)} & \textbf{5.32} \tiny{(-4.95\%)}   \\
	 \midrule
	 \textbf{$\alpha_m=0.3$} &&&   \\
	    & RNN-T             & 8.65  &  21.88  \\
	    & RNN-T (Ours)      & \textbf{8.21} \tiny{(-5.09\%)}  & \textbf{20.41} \tiny{(-6.72\%)}  \\
	    & Conformer         & 2.79  & 6.72   \\
	    & Conformer (Ours)  & \textbf{2.56} \tiny{(-8.24\%)}  & \textbf{5.86} \tiny{(-12.79\%)}  \\
	\bottomrule
\end{tabularx}
\vspace{-1.5em}
\end{table}

\autoref{tab:r1} demonstrates the performance of our model on both the RNN-T and Conformer encoders when trained and evaluated on the Librispeech dataset. We can see here that using the factored representation alone does not add much additional benefit, which as discussed in \autoref{subsec:factorized}, makes some sense as forcing the model to encode foreground and background separately does not actually add any representational power on its own. Additionally notable, adding background-contrastive learning to a non-factored representation actually performs worse in most situations, suggesting that alone (without the factorization), the contrastive learning only serves to confuse the model. Of the single techniques, only MLM improves performance. When combining the factored representation with the background-contrastive learning or MLM, performance notably improves, and combining all three of the losses leads to the optimal improvement over the baseline models. 

These gains are not only limited to Librispeech. \autoref{tab:r2} demonstrates the performance of our model when trained on real-world Alexa-AI benchmarks. From these results, we can see that the model achieves similar gains on both the Base and Rare datasets, with gains on the rare data word-error-rate slightly outpacing those on the base dataset. Because we expect our model to mainly handle noise (and not necessarily allow for additional representation), the fact that the gains are similar on these datasets makes some sense (as they are drawn from the same noise distributions). Notably, the model seems to make the most gains in substitutions and deletions, which are often caused by noisy situations.

To further validate the fact that our model is performing well in noisy scenarios, \autoref{tab:r4} demonstrates the performance of our method on artificially noisy data. We can see that while the performance degrades rapidly as we introduce noisy data (up to 30\%  mixed samples), our model is able to significantly outperform the baseline method. It's interesting to see that for both \autoref{tab:r4} and \autoref{tab:r1}, the performance on the test-other dataset improves more, relative to the test-clean dataset. This performance increase is likely due to the data: in the test-other dataset, modeling the background is far more important than modeling the foreground.
\section{Conclusion}

In this work, we have introduced a novel unsupervised, encoder-agnostic method for factoring speech-encoder representations into explicit content-encoding representations, and spurious context-encoding representations. We demonstrate that our method performs well on several encoders and datasets, as well as that our method is robust to both real-world and artificial noise. While this work is a strong first step towards exploring disentangled latent spaces for ASR, there is significant space to expand how we train such factored representations. While our current method factors along the explicit axes of content and context, we believe that the number of factors could be expanded to include variables such as visual and multi-modal referential content, speaker language and identity, device and input-specific profiles and geographic or regional factors among others. While such factors may require additional labeled data, they can help to further build disentangled representations of audio which focus on content over context. We strongly believe that such disentangled representations present a path forward for integrating and improving ASR models in a wide range of domains and applications. 
\clearpage

\section{References}
\begingroup
  \def\section*#1{}
  \small
  \setlength{\bibsep}{4pt}
  \bibliographystyle{IEEEtran}
  \bibliography{references}

\begin{thebibliography}{10}
\providecommand{\url}[1]{#1}
\csname url@samestyle\endcsname
\providecommand{\newblock}{\relax}
\providecommand{\bibinfo}[2]{#2}
\providecommand{\BIBentrySTDinterwordspacing}{\spaceskip=0pt\relax}
\providecommand{\BIBentryALTinterwordstretchfactor}{4}
\providecommand{\BIBentryALTinterwordspacing}{\spaceskip=\fontdimen2\font plus
\BIBentryALTinterwordstretchfactor\fontdimen3\font minus
  \fontdimen4\font\relax}
\providecommand{\BIBforeignlanguage}[2]{{%
\expandafter\ifx\csname l@#1\endcsname\relax
\typeout{** WARNING: IEEEtran.bst: No hyphenation pattern has been}%
\typeout{** loaded for the language `#1'. Using the pattern for}%
\typeout{** the default language instead.}%
\else
\language=\csname l@#1\endcsname
\fi
#2}}
\providecommand{\BIBdecl}{\relax}
\BIBdecl

\bibitem{shi2022learning}
B.~Shi, W.-N. Hsu, K.~Lakhotia, and A.~Mohamed, ``Learning audio-visual speech
  representation by masked multimodal cluster prediction,'' \emph{arXiv
  preprint arXiv:2201.02184}, 2022.

\bibitem{chan2021multi}
D.~M. Chan, S.~Ghosh, D.~Chakrabarty, and B.~Hoffmeister, ``Multi-modal
  pre-training for automated speech recognition,'' \emph{arXiv preprint
  arXiv:2110.09890}, 2021.

\bibitem{bertoin2021disentanglement}
D.~Bertoin and E.~Rachelson, ``Disentanglement by cyclic reconstruction,''
  \emph{arXiv preprint arXiv:2112.12980}, 2021.

\bibitem{kwon2020intra}
Y.~Kwon, S.-W. Chung, and H.-G. Kang, ``Intra-class variation reduction of
  speaker representation in disentanglement framework,'' \emph{arXiv preprint
  arXiv:2008.01348}, 2020.

\bibitem{du2021identity}
Z.~Du, B.~Sisman, K.~Zhou, and H.~Li, ``Identity conversion for emotional
  speakers: A study for disentanglement of emotion style and speaker
  identity,'' \emph{arXiv preprint arXiv:2110.10326}, 2021.

\bibitem{chen2021scenario}
S.-J. Chen, W.~Xia, and J.~H. Hansen, ``Scenario aware speech recognition:
  Advancements for apollo fearless steps \& chime-4 corpora,'' \emph{arXiv
  preprint arXiv:2109.11086}, 2021.

\bibitem{tenenbaum2000separating}
J.~B. Tenenbaum and W.~T. Freeman, ``Separating style and content with bilinear
  models,'' \emph{Neural computation}, vol.~12, no.~6, pp. 1247--1283, 2000.

\bibitem{rifai2012disentangling}
S.~Rifai, Y.~Bengio, A.~Courville, P.~Vincent, and M.~Mirza, ``Disentangling
  factors of variation for facial expression recognition,'' in \emph{European
  Conference on Computer Vision}.\hskip 1em plus 0.5em minus 0.4em\relax
  Springer, 2012, pp. 808--822.

\bibitem{mathieu2016disentangling}
M.~F. Mathieu, J.~J. Zhao, J.~Zhao, A.~Ramesh, P.~Sprechmann, and Y.~LeCun,
  ``Disentangling factors of variation in deep representation using adversarial
  training,'' \emph{Advances in neural information processing systems},
  vol.~29, 2016.

\bibitem{chen2016infogan}
X.~Chen, Y.~Duan, R.~Houthooft, J.~Schulman, I.~Sutskever, and P.~Abbeel,
  ``Infogan: Interpretable representation learning by information maximizing
  generative adversarial nets,'' \emph{Advances in neural information
  processing systems}, vol.~29, 2016.

\bibitem{higgins2016beta}
I.~Higgins, L.~Matthey, A.~Pal, C.~Burgess, X.~Glorot, M.~Botvinick,
  S.~Mohamed, and A.~Lerchner, ``beta-vae: Learning basic visual concepts with
  a constrained variational framework,'' 2016.

\bibitem{sanchez2020learning}
E.~H. Sanchez, M.~Serrurier, and M.~Ortner, ``Learning disentangled
  representations via mutual information estimation,'' in \emph{European
  Conference on Computer Vision}.\hskip 1em plus 0.5em minus 0.4em\relax
  Springer, 2020, pp. 205--221.

\bibitem{xin2021disentangled}
D.~Xin, T.~Komatsu, S.~Takamichi, and H.~Saruwatari, ``Disentangled speaker and
  language representations using mutual information minimization and domain
  adaptation for cross-lingual tts,'' in \emph{ICASSP 2021-2021 IEEE
  International Conference on Acoustics, Speech and Signal Processing
  (ICASSP)}.\hskip 1em plus 0.5em minus 0.4em\relax IEEE, 2021, pp. 6608--6612.

\bibitem{gong2021using}
C.~Gong, L.~Wang, Z.~Ling, J.~Zhang, and J.~Dang, ``Using multiple reference
  audios and style embedding constraints for speech synthesis,'' \emph{arXiv
  preprint arXiv:2110.04451}, 2021.

\bibitem{wang2021vqmivc}
D.~Wang, L.~Deng, Y.~T. Yeung, X.~Chen, X.~Liu, and H.~Meng, ``Vqmivc: Vector
  quantization and mutual information-based unsupervised speech representation
  disentanglement for one-shot voice conversion,'' \emph{arXiv preprint
  arXiv:2106.10132}, 2021.

\bibitem{devlin2018bert}
J.~Devlin, M.-W. Chang, K.~Lee, and K.~Toutanova, ``Bert: Pre-training of deep
  bidirectional transformers for language understanding,'' \emph{arXiv preprint
  arXiv:1810.04805}, 2018.

\bibitem{belghazi2018mine}
M.~I. Belghazi, A.~Baratin, S.~Rajeswar, S.~Ozair, Y.~Bengio, A.~Courville, and
  R.~D. Hjelm, ``Mine: mutual information neural estimation,'' \emph{arXiv
  preprint arXiv:1801.04062}, 2018.

\bibitem{ganin2016domain}
Y.~Ganin, E.~Ustinova, H.~Ajakan, P.~Germain, H.~Larochelle, F.~Laviolette,
  M.~Marchand, and V.~Lempitsky, ``Domain-adversarial training of neural
  networks,'' \emph{The journal of machine learning research}, vol.~17, no.~1,
  pp. 2096--2030, 2016.

\bibitem{chen2020simple}
T.~Chen, S.~Kornblith, M.~Norouzi, and G.~Hinton, ``A simple framework for
  contrastive learning of visual representations,'' in \emph{International
  conference on machine learning}.\hskip 1em plus 0.5em minus 0.4em\relax PMLR,
  2020, pp. 1597--1607.

\bibitem{baevski2020wav2vec}
A.~Baevski, Y.~Zhou, A.~Mohamed, and M.~Auli, ``wav2vec 2.0: A framework for
  self-supervised learning of speech representations,'' \emph{Advances in
  Neural Information Processing Systems}, vol.~33, pp. 12\,449--12\,460, 2020.

\bibitem{hsu2021hubert}
W.-N. Hsu, Y.-H.~H. Tsai, B.~Bolte, R.~Salakhutdinov, and A.~Mohamed, ``Hubert:
  How much can a bad teacher benefit asr pre-training?'' in \emph{ICASSP
  2021-2021 IEEE International Conference on Acoustics, Speech and Signal
  Processing (ICASSP)}.\hskip 1em plus 0.5em minus 0.4em\relax IEEE, 2021, pp.
  6533--6537.

\bibitem{graves2013speech}
A.~Graves, A.-r. Mohamed, and G.~Hinton, ``Speech recognition with deep
  recurrent neural networks,'' in \emph{2013 IEEE international conference on
  acoustics, speech and signal processing}.\hskip 1em plus 0.5em minus
  0.4em\relax Ieee, 2013, pp. 6645--6649.

\bibitem{gulati2020conformer}
A.~Gulati, J.~Qin, C.-C. Chiu, N.~Parmar, Y.~Zhang, J.~Yu, W.~Han, S.~Wang,
  Z.~Zhang, Y.~Wu \emph{et~al.}, ``Conformer: Convolution-augmented transformer
  for speech recognition,'' \emph{arXiv preprint arXiv:2005.08100}, 2020.

\bibitem{kudo2018sentencepiece}
T.~Kudo and J.~Richardson, ``Sentencepiece: A simple and language independent
  subword tokenizer and detokenizer for neural text processing,'' \emph{arXiv
  preprint arXiv:1808.06226}, 2018.

\bibitem{panayotov2015librispeech}
V.~Panayotov, G.~Chen, D.~Povey, and S.~Khudanpur, ``Librispeech: an asr corpus
  based on public domain audio books,'' in \emph{2015 IEEE international
  conference on acoustics, speech and signal processing (ICASSP)}.\hskip 1em
  plus 0.5em minus 0.4em\relax IEEE, 2015, pp. 5206--5210.

\bibitem{oppenheim2001discrete}
A.~V. Oppenheim, J.~R. Buck, and R.~W. Schafer, \emph{Discrete-time signal
  processing. Vol. 2}.\hskip 1em plus 0.5em minus 0.4em\relax Upper Saddle
  River, NJ: Prentice Hall, 2001.

\bibitem{park2019specaugment}
D.~S. Park, W.~Chan, Y.~Zhang, C.-C. Chiu, B.~Zoph, E.~D. Cubuk, and Q.~V. Le,
  ``Specaugment: A simple data augmentation method for automatic speech
  recognition,'' \emph{arXiv preprint arXiv:1904.08779}, 2019.

\end{thebibliography}
\endgroup
\end{document}